\documentclass[sn-mathphys,Numbered]{sn-jnl}

\usepackage{graphicx}%
\usepackage{multirow}%
\usepackage{amsmath,amssymb,amsfonts}%
\usepackage{amsthm}%
\usepackage{mathrsfs}%
\usepackage[title]{appendix}%
\usepackage{xcolor}%
\usepackage{textcomp}%
\usepackage{manyfoot}%
\usepackage{booktabs}%
\usepackage{algorithm}%
\usepackage{algorithmicx}%
\usepackage{algpseudocode}%
\usepackage{listings}%

\begin{document}

\title[Article Title]{Theoretical study of normal deformed rotational bands of odd-mass $^{167,169,171}$Lu nuclei}

\author*[1]{\fnm{Carlos E} \sur{Vargas}}\email{cavargas@uv.mx}

\author[2]{\fnm{V\'ictor} \sur{Vel\'azquez-Aguilar}}\email{vicvela@ciencias.unam.mx}

\author[1,3]{\fnm{Jes\'us} \sur{\'Avila-Pulido}}\email{jesus.avila-pulido@psi.ch}

\affil*[1]{\orgdiv{Facultad de F\'isica}, \orgname{Universidad Veracruzana}, \orgaddress{\street{Paseo No. 112, Des. Hab. Nvo. Xalapa}, \city{Xalapa}, \postcode{91097}, \state{Veracruz}, \country{Mexico}}}

\affil[2]{\orgdiv{Facultad de Ciencias}, \orgname{Universidad Nacional Aut\'onoma de M\'exico}, \orgaddress{\street{Apartado Postal 70-542},
\postcode{04510}, \state{Mexico City}, \country{Mexico}}}

\affil[3]{\orgdiv{}\orgname{Paul Scherrer Institut}, \orgaddress{\street{Forschungsstrasse 111},
\postcode{CH-5232}, \state{Villigen PSI}, \country{Switzerland}}}

\abstract{The theoretical description of nuclear structure in rare earth nuclei represents a significant challenge for many shell models. This difficulty stems from the size of the configuration space involved. Methods based on symmetries provide significant advantages. Furthermore, the pseudo-SU(3) shell model has proven to be very useful in the description of these systems. In particular, we propose to study the energy spectra of the rotational bands with normal deformation at low and medium spin ($J\leq35/2$), the quadrupole moments, and the deformation parameter $\gamma$ associated with the states, as well as the B(E2) transition strengths in the $^{167,169,171}$Lu nuclei. The Hamiltonian includes $Q \cdot Q$, Nilsson, and pairing terms, parameterized in a systematic way, in addition to three rotor-like terms that enable fine tuning of the spectra. Our calculations predict collective rotational bands with prolate normal deformation in most cases. The exception is the $1/2^+$ band in $^{171}$Lu with a triaxial shape. The quadrupole moments vary along the bands, increasing in absolute value with spin. The $\gamma$ deformation parameter remains nearly constant along the bands, and the B(E2) values suggest bands with collective character and normal deformation. The agreement and limitations of the model are discussed.}


\maketitle

\section{Introduction}\label{intro}

The complex interplay between collective and single particle degrees of freedom in nuclear physics is particularly fascinating. The rare earth region is characterized by the collective nature of its constituent nuclei \cite{Boh98}. These nuclei exhibit rotational bands that result from the collective rotation of many particles. However, at high angular momentum (spin) or in neutron-deficient isotopes, a smooth terminating band could occur, showing a continuous transition from high collectivity to a pure particle-hole state \cite{Afa99}. 

The neutron-deficient nuclei $^{167,169,171}$Lu are located in the rare-earth region, where the low-lying states are mainly highly prolate with permanent quadrupole deformation ($\epsilon \sim 0.250$) \cite{Mol95}. However, at high spin, band termination occurs due to the alignment of quasi-particles in orbitals with high-$j$. This alignment contributes to stabilising triaxial deformation and enables wobbling-type oscillations \cite{Ode01}. As initially theorised by Bohr and Mottelson \cite{Boh98}, these oscillations consist of the precession of the angular momentum vector along the principal axis of inertia. Several groups have studied the nature of the wobbling excitation mode in isotopes of lutetium isotopes, particularly in $^{167}$Lu \cite{Amr03,Rou15}, where the existence of this mode has been firmly established.

Based on the experimental observation of strongly deformed bands at high spin \cite{Sch92} and calculations performed with Cranker codes \cite{Ben89,Ben90}, the literature  commonly refers to these excited bands, produced by wobbling harmonic vibrations, as triaxial strongly deformed (TSD) bands \cite{Amr03}. However, this terminology has been debated \cite{Rag17} due to doubts about the existence of a secondary, very deformed minimum. The main arguments of the criticism are threefold: i) the observed values of the quadrupole moments indicate less deformation than expected, ii) the strong interaction of the normally deformed bands with those TSD-bands observed at high spin, and iii) the low dynamical moment of inertia, which implies less deformation than would be assumed for strong deformation. The triaxial character is clear, but additional elements are needed to determine whether these are truly strongly deformed bands. 

Detailed measurements and theoretical studies of quadrupole moments and energy level schemes could provide greater certainty about the structure of these bands and more confident interpretation of such excitations \cite{Rag17}. This work aims to describe normal deformed bands and their properties at low and medium spin, contributing to the discussion of collective intrinsic structures and the possible existence of triaxial states \cite{Bon25}. Couplings with quasiparticles in high-$j$ orbitals are not considered in this contribution; therefore, studying wobbling oscillations lies beyond our scope.

One of the preferred theoretical schemes for describing the microscopic properties of nuclei is the nuclear shell model. In 1958 \cite{Ell58,Ell582}, Elliott introduced the SU(3) model, based on symmetries, which has proven highly useful in describing light-deformed nuclei. However, in heavy nuclei, the strong spin-orbit interaction breaks the harmonic oscillator symmetry, rendering the SU(3) model inapplicable. Consequently, pseudo-spin symmetry \cite{Ari69,Hec69} with relativistic origins \cite{Gin77} appears. This refers to the experimental fact that single-particle orbitals with $j=l-1/2$ and $j=(l-2)+1/2$ in the shell $\eta$ lie very close in energy. This proximity allows them to be labelled as pseudo-spin doublets with quantum numbers $\tilde{j}=j$, $\tilde{\eta}=\eta-1$, and $\tilde{l}=l-1$. Based on this symmetry, the pseudo-SU(3) coupling scheme was developed \cite{Rat73}, proving useful in the microscopic description of heavy deformed nuclei. The first calculations treated the model as a dynamical symmetry, employing only the leading irreducible representation (irrep) to describe the yrast band \cite{Dra84,Cas87}. The development of a computational code \cite{Bah94} to calculate triply reduced matrix elements between different SU(3) irreps has allowed the consideration of more realistic Hamiltonians, with breaking symmetry terms such as the Nilsson single-particle energies and pairing correlations, and has enabled the study \cite{Var98} of the interplay between the quadrupole-quadrupole and spin-orbit interactions. The strongest limitation of the pseudo-SU(3) model should be noted: it is not able to describe the abnormal parity states because the intruder orbitals, which lie lower in energy due to the strong spin-orbit coupling, have been explicitly left out of the description.

To provide insights into determining the nature of rotational bands with normal deformation at low and medium spin, in this work, we employ the pseudo-SU(3) shell model, based on symmetries, to obtain a theoretical description of low-lying rotational bands for $J^\pi \leq \frac{31}{2}^+$ in $^{171}$Lu and $J^\pi \leq \frac{35}{2}^+$ in the $^{167,169}$Lu nuclei. The calculated energy values are reported and compared with available experimental data. Additionally, the quadrupole-moments and the deformation parameter $\gamma$ of the levels that form the rotational bands are presented, along with some intra- and inter-band B(E2) transition strengths between these bands. These calculations could be useful in determining the contribution of collective bands at medium or high spin to the wobbling oscillations.

In Section \ref{model} a brief description of the pseudo-SU(3) classification scheme, the Hamiltonian employed and the expression of the $\gamma$ deformation parameter are discussed. In Section \ref{sec-en}, we present results for energies of levels for ground-state, $1/2^+$ and $5/2^+$ bands in $^{171,169,167}$Lu nuclei. They are presented in order of decreasing stability: $^{171}$Lu, $^{169}$Lu, $^{167}$Lu, moving progressively towards the neutron drip line. Electric quadrupole moments, together with the deformation parameter $\gamma$ and intra- and inter-band B(E2) transitions in these nuclei are discussed in Section \ref{qm-gamma-be2}. Finally, a brief conclusion is given in Section \ref{conclu}.

\section{The model}\label{model}

The starting point in constructing a shell model is the selection of the basis for the many-body states. The pseudo-SU(3) model has been extensively discussed \cite{Tro95,Var00b}, including general expressions for the basis states and the matrix elements of generic operators between them. Hence, this Section provides only a brief summary of the most important elements of the model.

For $n_\alpha$ active nucleons ($\alpha = \pi, \nu$) in a shell of normal parity $\eta^N_\alpha$, the many-particle state is classified according to the following chain of groups \cite{Dra84,Cas87,Var00} $U(2\Omega^N_\alpha ) \supset U(\Omega^N_\alpha ) \times U(2) \supset SU(3) \times SU(2) \supset SO(3) \times SU(2) \supset SU_J(2)$, where the irreducible representations (irreps) of $U(2\Omega^N_\alpha)$ are labelled by $\{ 1^{n^{N}_\alpha}\}$, and the SU(3) representations are labelled by $(\lambda_\alpha,\mu_\alpha)$, with the multiplicity labels of the implied reductions ($\gamma_\alpha$ and $K_\alpha$) also being considered. Applying this framework to the lutetium isotopes, these nuclei are composed of 71 protons, 50 of which form the inner core ($\eta_\pi$ = 3), while 21 occupy orbitals in the $\eta_\pi$ = 4 shell. Of these, 13 are in normal parity levels, and 8 are in abnormal parity $h_{11/2}$ levels. For neutrons, 82 close the inner core ($\eta_\nu$ = 4), and the occupation numbers for the remaining neutrons in shell $\eta_\nu$ = 5 are presented in Table \ref{occup}.
\begin{table}
\begin{tabular}{ccccc}
&&&& \\ \hline \hline
     Nucleus        & $\epsilon_2$ & $n_\nu$ & $n_\nu^N$ & $n_\nu^A$ \\ \hline 
 $^{167}$Lu$_{96}$  &  0.250   &   14    &   10      &  4       \\
 $^{169}$Lu$_{98}$  &  0.258   &   16    &   10      &  6       \\
 $^{171}$Lu$_{100}$ &  0.258   &   18    &   12      &  6       \\ \hline \hline
\end{tabular}
\caption{Deformation ($\epsilon_2$) \cite{Mol95} and occupation numbers ($n$) for neutrons ($\nu$). The superscript $N$ and $A$ indicate normal and abnormal parity levels, respectively.}
\label{occup}
\end{table}

The pseudo-SU(3) model has been used to describe the rotational bands with normal parity and deformation and the intra- and inter-band B(E2) transitions in even-even and A-odd nuclei \cite{Var00,Var01,Dra04}. In such studies, the Hilbert space considered has been composed of states with normal parity $ | \beta J M \rangle $ with the maximum spatial symmetry $\tilde{S}_{\pi,\nu}$ = 0 (for an even) and 1/2 (for an odd) number of protons and neutrons. In more recent studies \cite{Var02,Var04,Hir06,Var13,Var17,Var24}, an extension of the Hilbert space has been used to include states with reduced spatial symmetry, ($\tilde{S}_{\pi,\nu}$ = 1 and 3/2, for an even and odd number of particles, respectively). This has allowed for the description of up to nine rotational bands in some nuclei, the analysis of the interplay between collective and single-particle degrees of freedom in the scissors mode in A-odd nuclei, and the description of spectra and electromagnetic properties in the isotopic chains $^{160-170}$Dy and $^{168-178}$Yb.

In the present work, the ground state and the excited $1/2^+$ and $5/2^+$ bands in the isotopes $^{167,169,171}$Lu at low and medium spin are described using this extension of the Hilbert space, which includes states with reduced spatial symmetry. As noted in Section \ref{intro}, nucleons in abnormal parity orbitals renormalise the dynamics described using nucleons in normal parity states \cite{Var98}. This limitation manifests itself in, for example, the need for high effective charges for describing quadrupole electromagnetic transitions. While this is the strongest limitation of the model, it has been demonstrated to be a reasonable approach \cite{Var17}.

The Hamiltonian considered in the present study is an extension of the many-body Nilsson Hamiltonian. In addition to the single-particle terms for protons and neutrons ($H_{sp,\pi[\nu]}$), it includes the quadrupole-quadrupole ($\tilde Q \cdot \tilde Q$) and pairing ($H_{pair,\pi[\nu]}$) interactions, as well as three rotor-like terms ($\tilde K^2$, $\tilde J^2$ and $\tilde C_3$) that are diagonal in the SU(3) basis:

\begin{eqnarray}
    \tilde H & = & \sum_{\alpha=\pi,\nu} \{ \tilde H_{sp,\alpha} - G_\alpha ~\tilde H_{pair,\alpha}	\} - \frac{1}{2}~  \chi~ \tilde Q \cdot \tilde Q \label{eq:ham} \\
      &   & + ~a~ \tilde K^2 + ~b~ \tilde J^2~ + ~c~ \tilde C_3. \nonumber
\end{eqnarray}

The basic components of any realistic Hamiltonian -- single-particle levels, pairing correlations, and the quadrupole-quadrupole interaction, essential for describing deformed nuclei -- are parameterised systematically as functions of mass $A$. Therefore, they are not treated as free parameters of the model \cite{Rin79,Duf96}. A detailed analysis of each term of the Hamiltonian and their parametrization can be found in Ref. \cite{Var00}. The mixing of SU(3) states is due to the presence of the symmetry-breaking Nilsson single-particle and pairing terms.

The second row of the Hamiltonian (\ref{eq:ham}) contains the rotor-like terms employed to fine-tune the spectra. Their three parameters $a$, $b$ and $c$ have been fitted following the methods presented in Refs. \cite{Var00,Var24}, where their effect on the energies has been discussed. For the isotopes studied ($^{171-167}$Lu), the $\tilde K^2$ term (breaking of the SU(3) degeneracy of the different $K$ bands \cite{Var00,Naq90}), parametrised by $a$, determines the band-head energy of the $1/2^+$ band. The term $\tilde J^2$ (which represents a small correction to the quadrupole–quadrupole term) provides additional moment of inertia to the rotational bands. Finally, the term $\tilde C_3$ (which is the Casimir invariant of third order) fits the band-head of the $5/2^+$ band. Table \ref{parame} shows the parameters used in the Hamiltonian (\ref{eq:ham}). It should be noted that the rotor-like terms help to fit the energies of rotational bands, while maintaining the wave functions of the states practically intact.
\begin{table}
\begin{tabular}{cccc}\hline \hline
Parameter& $^{171}$Lu & $^{169}$Lu & $^{167}$Lu \\ \hline
$a$	     &   48.00    &     67.00  &  84.05     \\   
$b$    	 &   -3.50    &     -3.10  &  -2.40     \\
$c$	     &    0.04    &     -0.03  &      0     \\ \hline \hline
\end{tabular}
\caption{Parameters (in keV) used in the Hamiltonian (\ref{eq:ham}).}
\label{parame}\end{table}

The electric quadrupole moment is a key observable for determining the deformation associated with the rotational bands, and it is crucial for understanding the triaxial nature of high spin excitations, as well as their connection with the states at intermediate spin ($J\sim 35/2$). Once the wave function has been obtained for each of the states that form the rotational bands, the electromagnetic properties can be evaluated. This involves calculating the expectation values of various electric and magnetic multipole operators. In previous works, the definitions for the probabilities of transition and electromagnetic moments have been presented \cite{Cas87,Var17} in detail. The interested reader is therefore referred to these works and the references therein.

To determine the deformation and triaxiality of rotational band states, we also present calculations of the expectation value of the $\gamma$ shape variable. In previous works, a connection has been established between the $\gamma$ and $\beta$ variables of the microscopic collective model (MCM) \cite{Row85}, which is an algebraic theory that has the dynamical group of the three-dimensional oscillator as its fundamental symmetry, and the irreducible representation (irrep) labels $\lambda$ and $\mu$ of SU(3) \cite{Cas88,Dra89}. These relationships are:
\begin{eqnarray}
\begin{aligned}
& \beta^2 =\frac{4\pi}{5} \frac{1}{(Ar_{rms}^2)^2}(\lambda^2 + 
  \lambda \mu + \mu^2 + 3\lambda + 3\mu + 3),\\ 
& \gamma = \mathrm{arctan} \left[ \frac{\sqrt{3}(\mu + 1)}{2 
  \lambda + \mu + 3} \right],
\label{eq:beta-gamma}
\end{aligned}
\end{eqnarray}

\noindent where $A$ is the total mass number of nucleons and $r_{rms}^2$ is the nuclear mean square radius. These equations allow us to associate each irrep, $(\lambda,\mu)$, with a specific point in the $(\beta,\gamma)$ plane.

\section{Low-lying energy spectra}\label{sec-en}

\begin{figure}[h]
\centering
\includegraphics[width=0.86\textwidth]{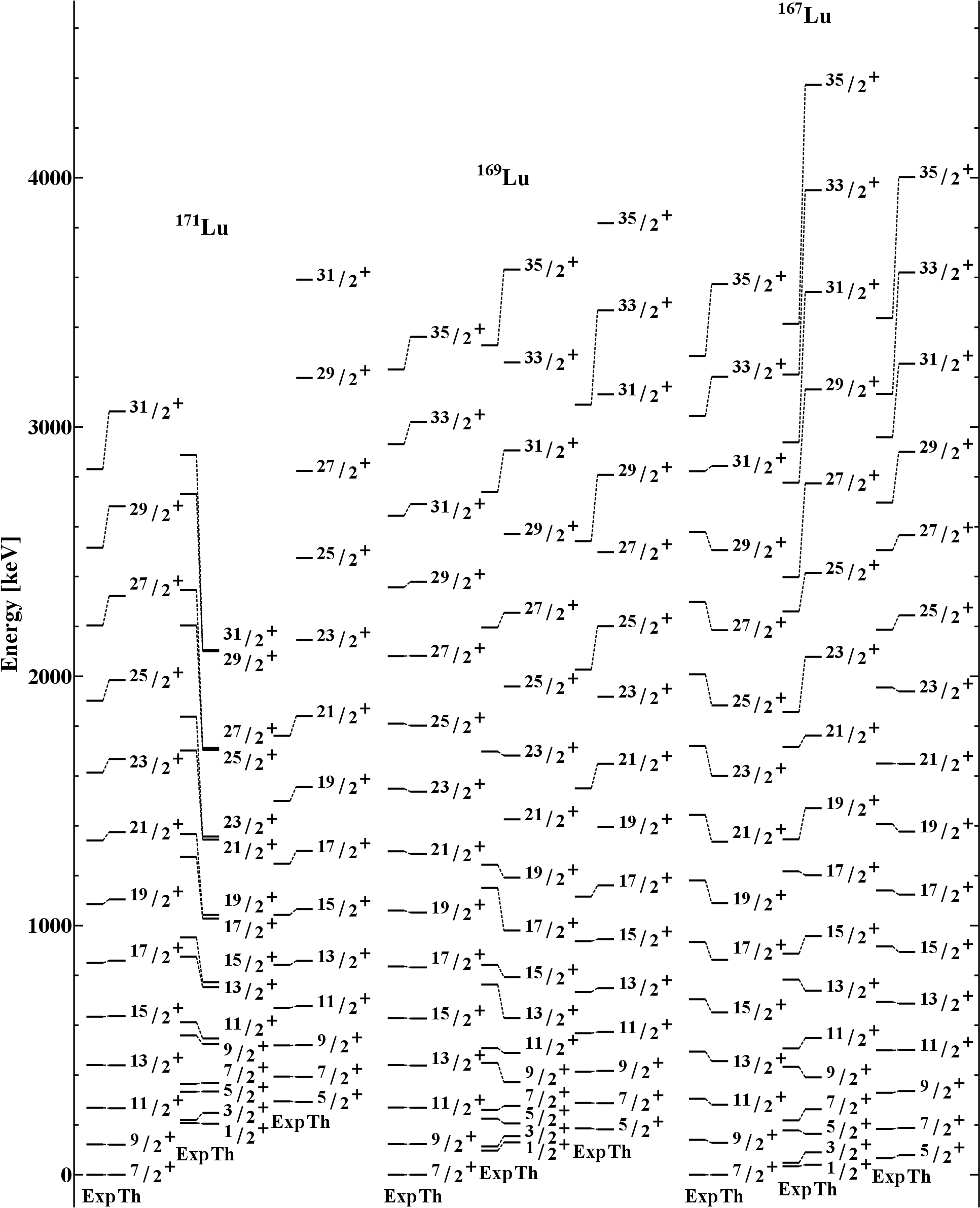}
\caption{Experimental and theoretical energies (in keV) of ground-state, $1/2^+$ and $5/2^+$ bands in $^{171-167}$Lu nuclei. The labels indicate the total angular momentum (spin) and parity of each level. Experimental data \cite{Sin23,Bag08,Bag18} are plotted on the left-hand side of each column, with theoretical values on the right. The correspondence between theoretical and experimental levels is indicated by dotted lines.}\label{energies}
\end{figure}

In this section, we present results for the energies of states in the ground-state band ($J^\pi=7/2^+$) and excited bands ($J^\pi=1/2^+$ and $5/2^+$) in the unstable isotopes $^{171}$Lu$_{100}$, $^{169}$Lu$_{98}$ and $^{167}$Lu$_{96}$. Fig. \ref{energies} shows the experimental and theoretical energies of these nuclei.

Previously, studies with the pseudo-SU(3) model in rare-earth nuclei (considering even-even and A-odd isotopes) have been conducted up to a maximum spin close to $J\sim10$ \cite{Var24}. In this work, for the first time, we have been able to employ the model to describe rotational bands up to medium spin ($J=35/2$), due to advances in computational power and memory capacity. For the states with $J=33/2$ and $35/2$ in the isotope $^{171}$Lu, the computational time needed to calculate the energies is prohibitively long; hence, it has not been possible to determine the properties of those states. Additionally, Table \ref{all-energies} presents the theoretical numerical values of the energies plotted in Fig. \ref{energies}, which may be useful for other research groups.
\begin{table}\hspace{-2.7cm}
\begin{tabular}{cccc|cccc|cccc}\hline \hline
 & \multicolumn{3}{c|}{Energies} & & \multicolumn{3}{c|}{Energies} & & \multicolumn{3}{c}{Energies} \\
$J_{band}^{\pi}$ & $^{171}$Lu & $^{169}$Lu & $^{167}$Lu & $J_{band}^{\pi}$ & $^{171}$Lu & $^{169}$Lu & $^{167}$Lu & $J_{band}^{\pi}$ & $^{171}$Lu & $^{169}$Lu & $^{167}$Lu \\ \hline
$7/2^+_{g.s.}$   &   0 &   0 &   0 &  $1/2^+_{1/2}$ & 205 & 130 &  40 &  $5/2^+_{5/2}$ & 292 & 182 &  78 \\
$9/2^+_{g.s.}$   & 121 & 123 & 128 &  $3/2^+_{1/2}$ & 249 & 155 &  90 &  $7/2^+_{5/2}$ & 393 & 287 & 188 \\
$11/2^+_{g.s.}$  & 267 & 269 & 281 &  $5/2^+_{1/2}$ & 334 & 206 & 164 &  $9/2^+_{5/2}$ & 520 & 417 & 336 \\
$13/2^+_{g.s.}$  & 439 & 438 & 456 &  $7/2^+_{1/2}$ & 369 & 276 & 263 & $11/2^+_{5/2}$ & 676 & 573 & 501 \\
$15/2^+_{g.s.}$  & 637 & 627 & 651 &  $9/2^+_{1/2}$ & 524 & 371 & 391 & $13/2^+_{5/2}$ & 858 & 749 & 687 \\
$17/2^+_{g.s.}$  & 859 & 832 & 862 & $11/2^+_{1/2}$ & 547 & 489 & 548 & $15/2^+_{5/2}$ &1066 & 945 & 894 \\
$19/2^+_{g.s.}$  &1105 &1052 &1090 & $13/2^+_{1/2}$ & 753 & 629 & 738 & $17/2^+_{5/2}$ &1299 &1161 &1124 \\
$21/2^+_{g.s.}$  &1375 &1287 &1336 & $15/2^+_{1/2}$ & 773 & 793 & 957 & $19/2^+_{5/2}$ &1557 &1396 &1377 \\
$23/2^+_{g.s.}$  &1668 &1537 &1600 & $17/2^+_{1/2}$ &1028 & 980 &1202 & $21/2^+_{5/2}$ &1840 &1649 &1649 \\
$25/2^+_{g.s.}$  &1984 &1802 &1883 & $19/2^+_{1/2}$ &1043 &1192 &1471 & $23/2^+_{5/2}$ &2145 &1918 &1939 \\
$27/2^+_{g.s.}$  &2322 &2082 &2185 & $21/2^+_{1/2}$ &1345 &1426 &1762 & $25/2^+_{5/2}$ &2474 &2201 &2244 \\
$29/2^+_{g.s.}$  &2682 &2379 &2506 & $23/2^+_{1/2}$ &1357 &1682 &2078 & $27/2^+_{5/2}$ &2824 &2498 &2566 \\
$31/2^+_{g.s.}$  &3063 &2691 &2844 & $25/2^+_{1/2}$ &1704 &1959 &2415 & $29/2^+_{5/2}$ &3197 &2808 &2901 \\
$33/2^+_{g.s.}$  & NA  &3020 &3202 & $27/2^+_{1/2}$ &1713 &2255 &2774 & $31/2^+_{5/2}$ &3591 &3131 &3254 \\
$35/2^+_{g.s.}$  & NA  &3362 &3574 & $29/2^+_{1/2}$ &2101 &2571 &3151 & $33/2^+_{5/2}$ & NA  &3468 &3620 \\
                 &     &     &     & $31/2^+_{1/2}$ &2106 &2906 &3542 & $35/2^+_{5/2}$ & NA  &3818 &4003 \\
                 &     &     &     & $33/2^+_{1/2}$ & NA  &3259 &3950 &                &     &     &     \\
                 &     &     &     & $35/2^+_{1/2}$ & NA  &3632 &4373 &                &     &     &     \\
\end{tabular}
\caption{Energies in $^{171,169,167}$Lu (given in keV).  Columns 1, 5, and 9 show the angular momentum (spin $J$) and parity ($\pi$) for the states in the ground-state, $1/2^+$ and $5/2^+$ bands, respectively. The remaining columns show theoretical energies from the pseudo-SU(3) model for $^{171}$Lu  (columns 2, 6 and 10), $^{169}$Lu (columns 3, 7 and 11) and $^{167}$Lu (columns 4, 8 and 12), respectively.}
\label{all-energies}\end{table}

In the ground-state band ($J^\pi=7/2^+$), the model predicts energies in good agreement with the experimental values in the three nuclei. However, at medium spin, the model underestimates the moment of inertia in the three nuclei (for example, in $^{167}$Lu for the state $J^\pi=35/2^+$, the difference $E^{Theo} - E^{Exp}$ is $289~keV$). 

For the rotational bands $1/2^+$, very strong staggering is observed in the experimental data in the three isotopes studied. The theoretical description of the staggering in these bands is incorrect in $^{167}$Lu and $^{169}$Lu; however, in $^{171}$Lu, the model overestimates the staggering compared to experimental data, resulting in nearly degenerate doublet energies. With respect to the theoretical values of the moment of inertia in these rotational bands, we find that in $^{171}$Lu, the $1/2^+$ band is predicted with an overestimated moment of inertia; in $^{169}$Lu, the moment of inertia is very similar to the experimental values; and in $^{167}$Lu, a rotational band with a very small moment of inertia is found. For the $5/2^+$ bands, the description of the energies is adequate, although in the three studied cases, a slight underestimate of the moment of inertia at medium spin is observed.

A common feature in the theoretical description of all rotational bands in all three isotopes is the lack of collectivity of the states when $J^\pi>29/2^+$. In almost all cases (except for the $1/2^+$ band in $^{171}$Lu), the model overestimates the energy of the excited states and underestimates the moment of inertia. The contribution of nucleons in intruder orbitals could explain this reduced collectivity in the states at medium spin, but, as previously pointed out, this sector has been explicitly excluded from the model; consequently, we cannot provide an explanation for this lack of collectivity in those states.

\section{Q-moments, $\gamma$ parameter and B(E2)s}\label{qm-gamma-be2}

In addition to the energies, we present calculated values for the electric quadrupole moments, the deformation parameter $\gamma$, and some intra- and inter-band B(E2) transition strengths. The electric quadrupole moment is calculated using the definition of the quadrupole transition operator following Refs. \cite{Var17,Cas87}. These references also provide the definitions for reduced matrix elements of SO(3) for the tensorial operator $T_{2\mu}$ between states of angular momentum $J_i$ and projection $M_i$. These observables provide insight into the nuclear structure and deformation properties.

The configuration space has been restricted to only one active shell; hence, core excitation effects, higher shell correlations, and polarization effects are not considered. This limitation of shell theories for nuclei is compensated for by using effective charges to calculate the quadrupole moments and B(E2) values. The effective charges used in the present work are $e_\pi = 2.3$ and $e_\nu = 1.3$. These are larger than those typically used in conventional shell model calculations due to the absence of nucleons in intruder levels. However, in previous works with the pseudo-SU(3) model \cite{Var17}, the same values for the effective charges have been used, enabling successful description of electric quadrupole moments and B(E2) values.

In Table \ref{qm}, the theoretical results for the electric quadrupole moments for each of the states of rotational bands in nuclei $^{171-167}$Lu are shown. 
\begin{table}\hspace{-2.7cm}
\begin{tabular}{cccc|cccc|cccc}\hline \hline
 & \multicolumn{3}{c|}{Q-moment [eb]} & & \multicolumn{3}{c|}{Q-moment [eb]} & & \multicolumn{3}{c}{Q-moment [eb]} \\
$J_{band}^{\pi}$ & $^{171}$Lu & $^{169}$Lu & $^{167}$Lu & $J_{band}^{\pi}$ & $^{171}$Lu & $^{169}$Lu & $^{167}$Lu & $J_{band}^{\pi}$ & $^{171}$Lu & $^{169}$Lu & $^{167}$Lu \\ \hline
$7/2^+_{g.s.}$   & 3.81& 3.60& 3.52&  $1/2^+_{1/2}$ &  NA &  NA &  NA &  $5/2^+_{5/2}$ & 2.93& 2.75& 2.75 \\
$9/2^+_{g.s.}$   & 1.46& 1.36& 1.27&  $3/2^+_{1/2}$ & 0.66&-1.11& 0.17&  $7/2^+_{5/2}$ & 0.58& 0.16&-0.26 \\
$11/2^+_{g.s.}$  & 0.05& 0.03&-0.06&  $5/2^+_{1/2}$ &-1.04&-1.81&-1.25&  $9/2^+_{5/2}$ &-0.51&-0.71&-0.75 \\
$13/2^+_{g.s.}$  &-0.86&-0.83&-0.94&  $7/2^+_{1/2}$ & 0.03&-1.92&-1.91& $11/2^+_{5/2}$ &-1.48&-1.46&-1.56 \\
$15/2^+_{g.s.}$  &-1.49&-1.42&-1.55&  $9/2^+_{1/2}$ &-0.22&-2.57&-2.23& $13/2^+_{5/2}$ &-1.99&-2.01&-2.16 \\
$17/2^+_{g.s.}$  &-1.94&-1.85&-1.99& $11/2^+_{1/2}$ & 0.16&-2.80&-2.34& $15/2^+_{5/2}$ &-2.36&-2.39&-2.50 \\
$19/2^+_{g.s.}$  &-2.27&-2.17&-2.32& $13/2^+_{1/2}$ & 0.20&-2.94&-2.36& $17/2^+_{5/2}$ &-2.63&-2.62&-2.67 \\
$21/2^+_{g.s.}$  &-2.54&-2.42&-2.53& $15/2^+_{1/2}$ & 0.28&-3.04&-2.48& $19/2^+_{5/2}$ &-2.83&-2.75&-2.76 \\
$23/2^+_{g.s.}$  &-2.74&-2.60&-2.67& $17/2^+_{1/2}$ & 0.30&-3.10&-2.63& $21/2^+_{5/2}$ &-2.96&-2.82&-2.81 \\
$25/2^+_{g.s.}$  &-2.90&-2.72&-2.77& $19/2^+_{1/2}$ & 0.36&-3.10&-2.81& $23/2^+_{5/2}$ &-3.12&-2.86&-2.85 \\
$27/2^+_{g.s.}$  &-3.04&-2.81&-2.84& $21/2^+_{1/2}$ & 0.37&-3.16&-2.95& $25/2^+_{5/2}$ &-3.22&-2.91&-2.88 \\
$29/2^+_{g.s.}$  &-3.15&-2.87&-2.88& $23/2^+_{1/2}$ & 0.40&-3.19&-3.08& $27/2^+_{5/2}$ &-3.31&-2.94&-2.93 \\
$31/2^+_{g.s.}$  &-3.19&-2.92&-2.92& $25/2^+_{1/2}$ & 0.40&-3.22&-3.17& $29/2^+_{5/2}$ &-3.39&-2.98&-2.95 \\
$33/2^+_{g.s.}$  &  NA &-2.95&-2.94& $27/2^+_{1/2}$ & 0.44&-3.25&-3.26& $31/2^+_{5/2}$ &-3.41&-3.01&-2.98 \\
$35/2^+_{g.s.}$  &  NA &-2.97&-2.96& $29/2^+_{1/2}$ & 0.44&-3.29&-3.31& $33/2^+_{5/2}$ &  NA &-3.03&-3.00 \\
                 &     &     &     & $31/2^+_{1/2}$ & 0.45&-3.32&-3.36& $35/2^+_{5/2}$ &  NA &-3.05&-3.03 \\
                 &     &     &     & $33/2^+_{1/2}$ &  NA &-3.34&-3.38&                &     &     &      \\
                 &     &     &     & $35/2^+_{1/2}$ &  NA &-3.36&-3.40&                &     &     &      \\
\end{tabular}
\caption{Theoretical electric quadrupole moments (given in $eb$) for rotational band states in $^{171,169,167}$Lu calculated using the pseudo-SU(3) model. Column layout follows Table \ref{all-energies}.}
\label{qm}\end{table}
Table \ref{qm} shows that the absolute value of the electric quadrupole moment increases to values close to $3 eb$ as the spin increases in the ground-state and $5/2^+$ bands, and to values close to $3.5 eb$ in the $1/2^+$ band. The exception is the $1/2^+$ band in $^{171}$Lu, where the quadrupole moment reaches a maximum of $0.45 eb$. For this nucleus, we could not calculate the quadrupole moments for values $J > 31/2$ due to prohibitively long computational times.

\begin{figure}[h]
\centering
\includegraphics[width=0.82\textwidth]{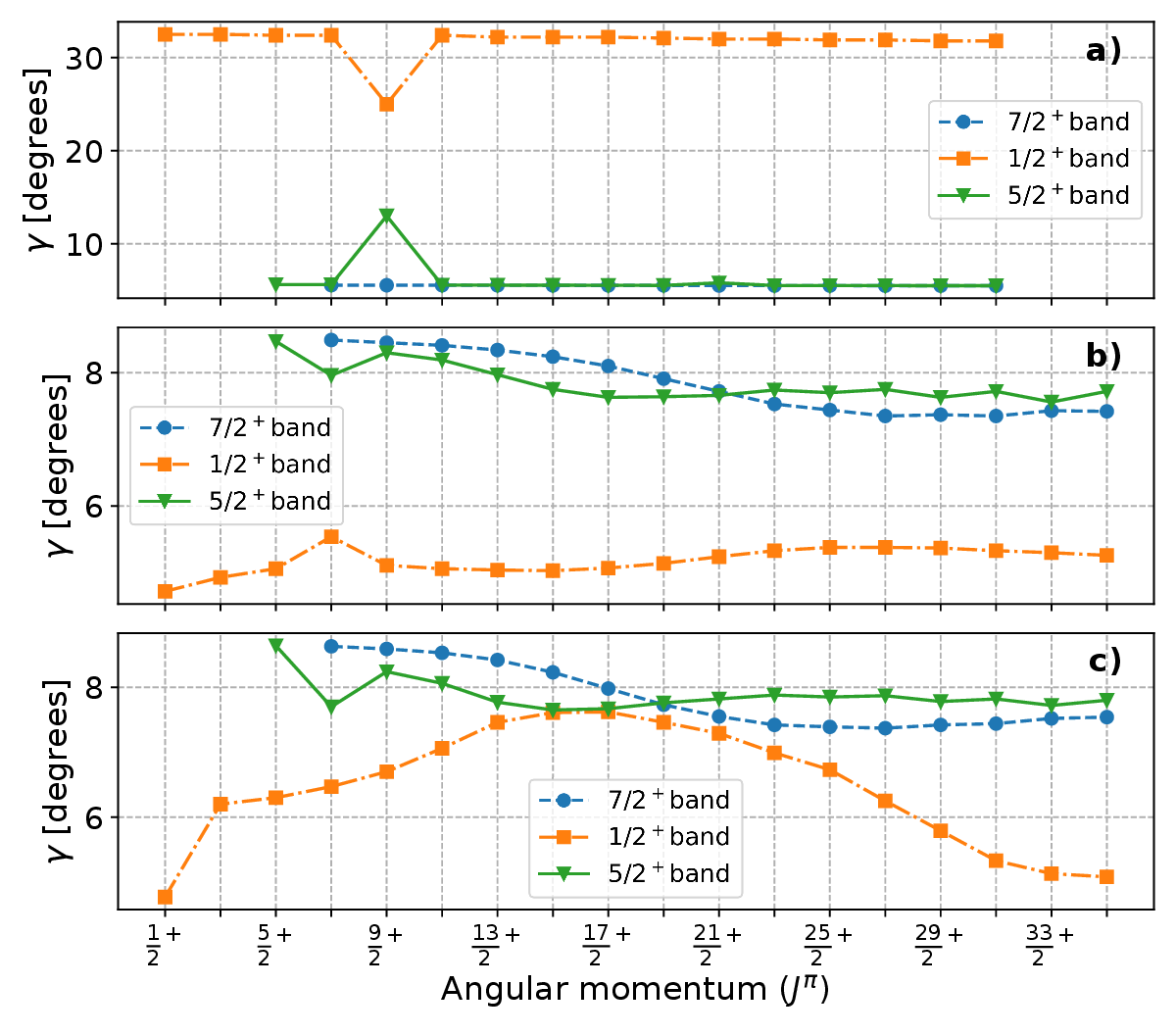}
\caption{Theoretical values of the $\gamma$ variable (in degrees) plotted against $J$ for: a) $^{171}$Lu, b) $^{169}$Lu, and c) $^{167}$Lu. In each graph, the line with circles represent the ground-state band, squares represent the $1/2^+$ band, and triangles represent the $5/2^+$ band.}\label{gamma-var}
\end{figure}

Using the wave function of the pseudo-SU(3) model and Equation (\ref{eq:beta-gamma}), we calculated the deformation parameter $\gamma$, see Fig. \ref{gamma-var}. 
The calculated values with the model show that, in the ground-state and $5/2^+$ bands, the deformation parameter $\gamma$ lies between $6^o$ and $8^o$. However, in the $^{171}$Lu nucleus, the $1/2^+$-band shows $\gamma \sim 32^o $, which is consistent with a strongly triaxial rotational band with normal deformation. Analyzing the wave function for the states of this band, we find that although the Nilsson energies and pairing terms of the Hamiltonian (\ref{eq:ham}) break the symmetry, mixing different pseudo-SU(3) states, the rotational $1/2^+$ band in $^{171}$Lu is formed almost exclusively (98.5\%) by the irrep $(\lambda,\mu)=(18,21)$, which corresponds to a triaxially deformed state.

In insert a) of Fig. \ref{gamma-var} ($^{171}$Lu), changes in the $\gamma$ parameter are observed at $J = 9/2^+$ in the $1/2^+$ and $5/2^+$ bands. This occurs due to the proximity of their energies ($E_{9/2^+} = 524$ keV for the $1/2^+$ band and $E_{9/2^+} = 520$ keV for the $5/2^+$ band; see Table \ref{all-energies}), resulting in probable state mixing.

In Table \ref{be2}, the intensities of the electric quadrupole transitions between some states of the calculated bands are shown. 
\begin{table}\hspace{-3.2cm}
\begin{tabular}{c|c|c|c}\hline \hline
 & \multicolumn{3}{c}{B(E2) [$e^2b^2 \times 10^{-2}$]}\\
$J_{i,band}^{\pi} \rightarrow J_{f,band}^{\pi}$ & $^{171}$Lu & $^{169}$Lu & $^{167}$Lu \\ \hline
%                 		     & Exp.&Theo.&   Exp.    &Theo.&     Exp.   &Theo.&    Exp.    &Theo.&    Exp.    &Theo.& Theo.\\ \hline
% ground state band transitions  J --> J + 1
$7/2^+_{g.s.}  \rightarrow 9/2^+_{g.s.}$  & 282 & 252 & 247 \\
$9/2^+_{g.s.}  \rightarrow 11/2^+_{g.s.}$ & 271 & 238 & 228 \\
$11/2^+_{g.s.} \rightarrow 13/2^+_{g.s.}$ & 224 & 191 & 176 \\
$13/2^+_{g.s.} \rightarrow 15/2^+_{g.s.}$ & 181 & 147 & 127 \\
%$15/2^+_{g.s.} \rightarrow 17/2^+_{g.s.}$ & 147 & 110 &  89 \\
% ground state band transitions  J --> J + 2
$7/2^+_{g.s.} \rightarrow 11/2^+_{g.s.}$  &  72 &  67 &  68 \\
$9/2^+_{g.s.} \rightarrow 13/2^+_{g.s.}$  & 124 & 115 & 117 \\
$11/2^+_{g.s.} \rightarrow 15/2^+_{g.s.}$ & 157 & 146 & 148 \\
$13/2^+_{g.s.} \rightarrow 17/2^+_{g.s.}$ & 178 & 166 & 170 \\ \hline
%$15/2^+_{g.s.} \rightarrow 19/2^+_{g.s.}$ & 191 & 180 & 187 \\ \hline
% 1/2-band transitions  J --> J + 1
$1/2^+_{1/2^+} \rightarrow 3/2^+_{1/2^+}$   & 268 & 206 & 106 \\
$3/2^+_{1/2^+} \rightarrow 5/2^+_{1/2^+}$   &$<$1 &  80 & 168 \\
$5/2^+_{1/2^+} \rightarrow 7/2^+_{1/2^+}$   &  26 &  34 &  94 \\
$7/2^+_{1/2^+} \rightarrow 9/2^+_{1/2^+}$   &   5 &  34 &  65 \\
% 1/2-band transitions  J --> J + 2
$1/2^+_{1/2^+} \rightarrow 5/2^+_{1/2^+}$   & 338 & 278 & 152 \\
$3/2^+_{1/2^+} \rightarrow 7/2^+_{1/2^+}$   & 305 & 232 & 213 \\
$5/2^+_{1/2^+} \rightarrow 9/2^+_{1/2^+}$   & 212 & 251 & 218 \\
$7/2^+_{1/2^+} \rightarrow 11/2^+_{1/2^+}$  & 316 & 231 & 196 \\ \hline
% 5/2-band transitions  J --> J + 2
$5/2^+_{5/2^+} \rightarrow 7/2^+_{5/2^+}$   & 315 & 233 & 198 \\
$7/2^+_{5/2^+} \rightarrow 9/2^+_{5/2^+}$   & 186 & 191 & 127 \\
$9/2^+_{5/2^+} \rightarrow 11/2^+_{5/2^+}$  & 135 & 153 & 133 \\
$11/2^+_{5/2^+} \rightarrow 13/2^+_{5/2^+}$ & 142 & 105 &  86 \\
% 5/2-band transitions  J --> J + 2
$5/2^+_{5/2^+} \rightarrow 9/2^+_{5/2^+}$   &  82 &  90 &  81 \\
$7/2^+_{5/2^+} \rightarrow 11/2^+_{5/2^+}$  & 164 & 139 &  71 \\
$9/2^+_{5/2^+} \rightarrow 13/2^+_{5/2^+}$  & 135 & 170 & 153 \\
$11/2^+_{5/2^+} \rightarrow 15/2^+_{5/2^+}$ & 208 & 185 & 183 \\ \hline \hline
% inter-band transitions
$3/2^+_{1/2^+} \rightarrow 5/2^+_{5/2^+}$   & $<$1&   1 &  1  \\
$5/2^+_{5/2^+} \rightarrow 7/2^+_{g.s.}$    &   3 &   4 & 10  \\
$5/2^+_{5/2^+} \rightarrow 9/2^+_{g.s.}$    &   2 &   2 &  5  \\
$3/2^+_{1/2^+} \rightarrow 7/2^+_{g.s.}$    & $<$1&   1 &  3  \\
$7/2^+_{g.s.} \rightarrow 7/2^+_{5/2^+}$    &   2 &   3 &  3  \\ \hline \hline
\end{tabular}
\caption{Intra- and inter-band B(E2;$J^+_{i,band} \rightarrow J^+_{f,band}$) transition strengths in $^{171-167}$Lu nuclei (given in $e^2b^2 \times 10^{-2}$). The first column shows the initial ($J_{i,band}$) and final ($J_{f,band}$) angular momentum (spin) values and bands. Columns 2-4 present theoretical pseudo-SU(3) model calculations for $^{171}$Lu, $^{169}$Lu, and $^{167}$Lu, respectively. Effective charges are $e_\pi = 2.3$ and $e_\nu = 1.3$.}
\label{be2}\end{table}
For the intra-band B(E2) transitions, we find that almost all the intensities are consistent with highly collective rotational bands, which connect states with similar wave functions. Particularly striking is the intra-band transition B(E2; $3/2^+_{1/2^+} \rightarrow 5/2^+_{1/2^+}$) in $^{171}$Lu, with an intensity less than $1 \times 10^{-2} e^2b^2$. This is due to mixing with other excited $5/2^+$ states at higher energies (not included in this work), whose wave functions also have a dominant triaxial component, distributing the total intensity among several transitions. Inter-band transitions are also included at the end of Table \ref{be2} for their possible utility in determining the triaxiality associated with the rotational bands.

\section{Conclusions}\label{conclu}

Motivated by community interest in understanding the nature and deformation of high-spin excited states in the rare-earth region, we have studied positive-parity bands with normal deformation in $^{171,169,167}$Lu isotopes using the pseudo-SU(3) shell model with symmetry-breaking terms, which has been previously used in the study of structure and electromagnetic properties of other nuclei in the rare-earth region. As in previous applications, the exclusion of the intruder sector remains the main limitation of the model.

Our results show that the model can describe the energy of the rotational bands at intermediate spins $J \sim 35/2$, much higher than in previous applications. This represents a significant achievement with the model and establishes the groundwork for extending the description to higher spins. In some cases, such as in $^{167}$Lu, the energies of the excited states at medium-spin ($J \sim 27/2$) of the $1/2^+$ and $5/2^+$ bands are overestimated. However, the triaxial $1/2^+$ band in $^{171}$Lu is exceptional, showing a very large moment of inertia and staggering, while the quadrupole moments of its constituent states are small. For this band, the wave function consists of states with very similar $\lambda$ and $\mu$ values, indicating a rotational band of collective character with a marked triaxial deformation. To summarise, the model demonstrates good predictive accuracy for energies. Some bands are almost perfectly described whilst others, such as the $1/2^+$ band in $^{171}$Lu and $^{167}$Lu, show incorrect energy predictions at medium spin.
 
In summary, our analysis indicates that the low and medium spin behavior in the $^{171-167}$Lu nuclei is consistent with prolate rotation developing some degree of collective triaxiality. This work provides a microscopic description of collective bands that could serve as a foundation for future investigations. High spin studies, considering the coupling of the collective states with quasiparticles in high-$j$ orbitals, represent a promising direction for future research with the model.

No funding was received for conducting this study. The authors have no competing interests which are relevant to the content of this article to declare.

%\section{Acknowledgments}


\begin{thebibliography}{1}

\bibitem{Boh98} Aage Bohr and Ben R. Mottelson. Nuclear Structure, Vol. I and II, (World Scientific Publishing Co, Singapore, 1998), Pags. 189-197 Vol. I and Pags. 22-44 Vol. II.
\bibitem{Afa99} A. V. Afanasjev, D. B. Fossan, G. J. Lane and I. Ragnarsson, Phys. Rep. 322, 1 (1999). DOI: https://doi.org/10.1016/S0370-1573(99)00035-6
\bibitem{Mol95} P. Möller, J.R. Nix, W.D. Myers, and W.J. Swiatecki. At. Data Nucl. Data Tables 59, 185 (1995). DOI: https://doi.org/10.1006/adnd.1995.1002
\bibitem{Ode01} S. W. Odegard {\it et al.}, Phys. Rev. Lett. 86, 5866 (2001). DOI: https://doi.org/10.1103/PhysRevLett.86.5866
\bibitem{Amr03} H. Amro {\it et al.}, Phys. Lett. B 553, 197 (2003). DOI: https://doi.org/10.1016/S0370-2693(02)03199-4
\bibitem{Rou15} D. G. Roux {\it et al.}, Phys. Rev. C 92, 064313 (2015). DOI: https://doi.org/10.1103/PhysRevC.92.064313
\bibitem{Sch92} W. Schmitz {\it et al.}, Nucl. Phys. A 539, 112 (1992). DOI: https://doi.org/10.1016/0375-9474(92)90238-F
\bibitem{Ben89} T. Bengtsson, Nucl. Phys. A 496, 56 (1989). DOI: https://doi.org/10.1016/0375-9474(89)90216-9
\bibitem{Ben90} T. Bengtsson, Nucl. Phys. A 512, 124 (1990). DOI: https://doi.org/10.1016/0375-9474(90)90007-9
\bibitem{Rag17} I. Ragnarsson, Phys. Scr. 92, 124004 (2017). DOI: 10.1088/1402-4896/aa9353
\bibitem{Bon25} D. Bonatsos {\it et al.}, 13, 47 (2025). DOI: https://doi.org/10.3390/atoms13060047
\bibitem{Ell58} J. P. Elliott, Proc. R. Soc. London, Ser. A 245, 128 (1958). DOI: http://doi.org/10.1098/rspa.1958.0072
\bibitem{Ell582} J. P. Elliott, Proc. R. Soc. London, Ser. A 245, 562 (1958). DOI: http://doi.org/10.1098/rspa.1958.0101
\bibitem{Ari69} A. Arima, M. Harvey and K. Shimizu, Phys. Lett. 30B, 517 (1969). DOI: https://doi.org/10.1016/0370-2693(69)90443-2
\bibitem{Hec69} K. T. Hecht and A. Adler, Nucl. Phys. A137, 129 (1969). DOI: https://doi.org/10.1016/0375-9474(69)90077-3
\bibitem{Gin77} J. N. Ginocchio, Phys. Rev. Lett. 78, 436 (1997). DOI: https://doi.org/10.1103/PhysRevLett.78.436
\bibitem{Rat73} R. D. Ratna-Raju, J. P. Draayer and K. T. Hecht, Nucl. Phys. A202 433 (1973). DOI: https://doi.org/10.1016/0375-9474(73)90635-0
\bibitem{Dra84} J. P. Draayer and K. J. Weeks, Ann. Phys. (N.Y.) 156, 41 (1984). DOI: https://doi.org/10.1016/0003-4916(84)90210-0
\bibitem{Cas87} O. Castaños, J. P. Draayer, and Y. Leschber, Ann. Phys. (N.Y.) 180, 290 (1987). DOI: https://doi.org/10.1016/0003-4916(87)90047-9
\bibitem{Bah94} C. Bahri and J. P. Draayer, Comput. Phys. Commun. 83, 59 (1994). DOI: https://doi.org/10.1016/0010-4655(94)90035-3
\bibitem{Var98} C. Vargas, J. G. Hirsch, P. O. Hess, and J. P. Draayer, Phys. Rev. C 58, 1488 (1998). DOI: https://doi.org/10.1103/PhysRevC.58.1488
\bibitem{Tro95} D. Troltenier, C. Bahri and J. P. Draayer, Nucl. Phys. A 586, 53 (1995). DOI: https://doi.org/10.1016/0375-9474(94)00518-R
\bibitem{Var00b} C. E. Vargas, J. G. Hirsch, T. Beuschel and J. P. Draayer, Phys. Rev. C 61, 031301(R) (2000). DOI: https://doi.org/10.1103/PhysRevC.61.031301
\bibitem{Var00} C. E. Vargas, J. G. Hirsch and J. P. Draayer, Nucl. Phys. A 673, 219 (2000). DOI: https://doi.org/10.1016/S0375-9474(00)00153-6
\bibitem{Var01} C. E. Vargas, J. G. Hirsch and J. P. Draayer, Phys. Rev. C 64, 034306 (2001). DOI: https://doi.org/10.1103/PhysRevC.64.034306
\bibitem{Dra04} J. P. Draayer, G. Popa, J. G. Hirsch, and  C. E. Vargas, High Energy Phys. Nucl. Phys. 28, 1297 (2004). DOI: https://hepnp.ihep.ac.cn/article/id/f8c2b816-6bfa-4d93-98f7-f4a5d13a705e
\bibitem{Var02} C. E. Vargas, J. G. Hirsch, and J. P. Draayer, Phys. Rev. C {\bf 66}, 064309 (2002). DOI: https://doi.org/10.1103/PhysRevC.66.064309
\bibitem{Var04} C. E. Vargas, and J. G. Hirsch, Phys. Rev. C 70, 064320 (2004). DOI: https://doi.org/10.1103/PhysRevC.70.064320
\bibitem{Hir06} J. G. Hirsch, G. Popa, S. R. Lesher, A. Aprahamian, C. E. Vargas, and J. P. Draayer, Rev. Mex. Fis. S {\bf 52}, 69 (2006).
\bibitem{Var13} C. E. Vargas, V. Velázquez and S. Lerma, Eur. Phys. J. A 49, 4 (2013). DOI: https://doi.org/10.1140/epja/i2013-13004-1
\bibitem{Var17} C. E. Vargas, V. Velázquez, S. Lerma and N. Bagatella, Eur. Phys. J. A 53, 73 (2017). DOI: https://doi.org/10.1140/epja/i2017-12264-y
\bibitem{Var24} C. E. Vargas and V. Velázquez-Aguilar, Eur. Phys. J. A 60, 138 (2024). DOI: https://doi.org/10.1140/epja/s10050-024-01354-y
\bibitem{Rin79} P. Ring and P. Schuck. {\it The Nuclear Many-Body Problem}, (Springer, Berlin, 1980), pp. 65-83.
\bibitem{Duf96} M. Dufour, A. P. Zuker, Phys. Rev. {\bf C 54} 1641 (1996). DOI: https://doi.org/10.1103/PhysRevC.54.1641
\bibitem{Naq90} H. A. Naqvi, J. P. Draayer, Nucl. Phys. A 516, 351 (1990). DOI: https://doi.org/10.1016/0375-9474(90)90313-B
\bibitem{Row85} D. J. Rowe, Rep. Prog. Phys. 48, 1419 (1985). DOI: 10.1088/0034-4885/48/10/003
\bibitem{Cas88} O. Castaños, J.P. Draayer, and Y. Leschber, Z. Phys. A 329, 33 (1988). DOI: https://doi.org/10.1007/BF01294813
\bibitem{Dra89} J. P. Draayer, S. C. Park, O. Casta\~nos, Phys. Rev. Lett. 62, 20 (1989). DOI: https://doi.org/10.1103/PhysRevLett.62.20
\bibitem{Sin23} B. Singh and J. Chen, Nucl. Data Sheets, 191, 1 (2023). DOI: https://doi.org/10.1016/j.nds.2023.08.001
\bibitem{Bag08} C. M. Baglin, Nucl. Data Sheets, 109, 2033 (2008). DOI: https://doi.org/10.1016/j.nds.2008.08.001
\bibitem{Bag18} C. M. Baglin and E. A. McCutchan, Nucl. Data Sheets, 151, 334 (2018). DOI: https://doi.org/10.1016/j.nds.2018.08.002

\end{thebibliography}
\end{document}